\documentclass[sigconf]{acmart}

\AtBeginDocument{%
  \providecommand\BibTeX{{%
    \normalfont B\kern-0.5em{\scshape i\kern-0.25em b}\kern-0.8em\TeX}}}

\setcopyright{acmlicensed}
\copyrightyear{2018}
\acmYear{2018}
\acmDOI{XXXXXXX.XXXXXXX}

\acmConference[Conference acronym 'XX]{Make sure to enter the correct
  conference title from your rights confirmation email}{June 03--05,
  2018}{Woodstock, NY}
\acmISBN{978-1-4503-XXXX-X/18/06}

\usepackage{multirow}
\begin{document}

\title{Evidence-Based Temporal  Fact Verification}


\author{Anab Maulana Barik \hspace*{0.3in} 
	Wynne Hsu \hspace*{0.3in}  Mong Li Lee}
\affiliation{National University of Singapore
\country{Singapore}}
\email{anabmaulana@u.nus.edu;  {whsu,leeml}@comp.nus.edu.sg}

\renewcommand{\shortauthors}{Barik et al}

\begin{abstract}
  Automated fact verification plays an essential role  in fostering trust in the digital space. Despite the growing interest, the verification of temporal facts  has not received much attention in the community. Temporal fact verification brings new challenges
 where cues of the temporal information need to be extracted and temporal reasoning involving various temporal aspects of the text must be applied.
In this work, we propose an end-to-end solution  for  temporal fact verification that considers the  temporal information in claims to obtain relevant evidence sentences and harness the power of large language model for temporal reasoning. Recognizing 
that  temporal facts often involve events, 
we model these events in the claim and evidence sentences.
We curate two temporal fact datasets to learn time-sensitive representations that encapsulate not only the semantic relationships among the events, but also their chronological proximity.
This allows us to retrieve the top-k
relevant evidence sentences and provide the context for a large language model to perform temporal reasoning and outputs whether a claim is supported or refuted by the retrieved evidence sentences.
Experiment results  demonstrate that the proposed approach significantly enhances the  accuracy  of temporal claim verification, thereby advancing current state-of-the-art in automated fact verification.
\end{abstract}





\maketitle

\section{Introduction}
The proliferation of false information, or "fake news," continues to pose a challenge  with potentially severe implications. Computational fact checking has been proposed as a viable solution to this issue, leveraging technology to verify textual claims against a set of evidence sentences that either support or contradict these claims. However, there is still a considerable gap when it comes to verifying temporal claims which are 
  statements associated with a specific time or duration. 
For effective verification of temporal claims, we need to retrieve evidences that focus not just on the semantic coherence between 
the claim and potential evidence, but more importantly, the temporal context so that the timeline is aligned between the claim and the evidences. 

 Consider the temporal claim \textit{"Matteo Renzi was a full-time undergraduate student in Singapore in 2006"}. This claim can be refuted if we find  evidence like "Matteo Renzi served as President of the Province of Florence from 2004 to 2009..." since it is highly unlikely for someone to serve as a president while concurrently undertaking a full-time undergraduate degree in a different country. Existing claim verification methods that employ traditional
evidence retrieval  based on lexical  or semantic matching
might overlook this evidence sentence and
 conclude that there is NOT ENOUGH INFO (NEI) to verify the claim.

Consider another  temporal claim \textit{"Henry Condell published his First Folio in 1623 and performed several plays for his career in 1620."}. We see that this claim has two events "\textit{published his First Folio}" and "\textit{performed several plays}" which are associated with 
two distinct dates "1623" and "1620" respectively.
For the temporal claim to be true, we need to verify that both events are supported by evidences sentences. 
On the other hand, if we have evidence that shows one of the event is false, then the entire claim becomes false.
For example, if we have the evidence 
sentence \textit{"Henry Condell ended his stage career in 1619."}, then we can 
refute the event that he performs several plays  in 1620, and conclude that the temporal claim is false.
By analyzing the claim and evidence sentence at the event-level rather than the whole sentence, we can 
link the time references to their respective events and retrieve relevant evidence sentences.

 
 Our work aims to bridge the gaps in temporal fact verification by introducing a framework called TACV that is sensitive to temporal aspects of claim  verification.
TACV  takes into account the temporal information in the claim to retrieve relevant evidence sentences.
 It  identifies events in
 both the claim and evidence sentences and associate the time-related information to the corresponding events.
With this, TACV seeks to assign a higher score to evidence sentences that align more closely with the events in the claim.
This is achieved through an encoder module  to create  temporal-aware representations. This process involves   
 modeling the events and their associated temporal information as a graph and  propagating the information using a graph attention network. The
representations obtained encapsulate  the temporal sequence and duration between dates of events, enabling us to 
account for the temporal disparity between the claim and the evidence sentences.

\begin{figure*}[t!] 
\centering
\includegraphics[width=0.82\textwidth]{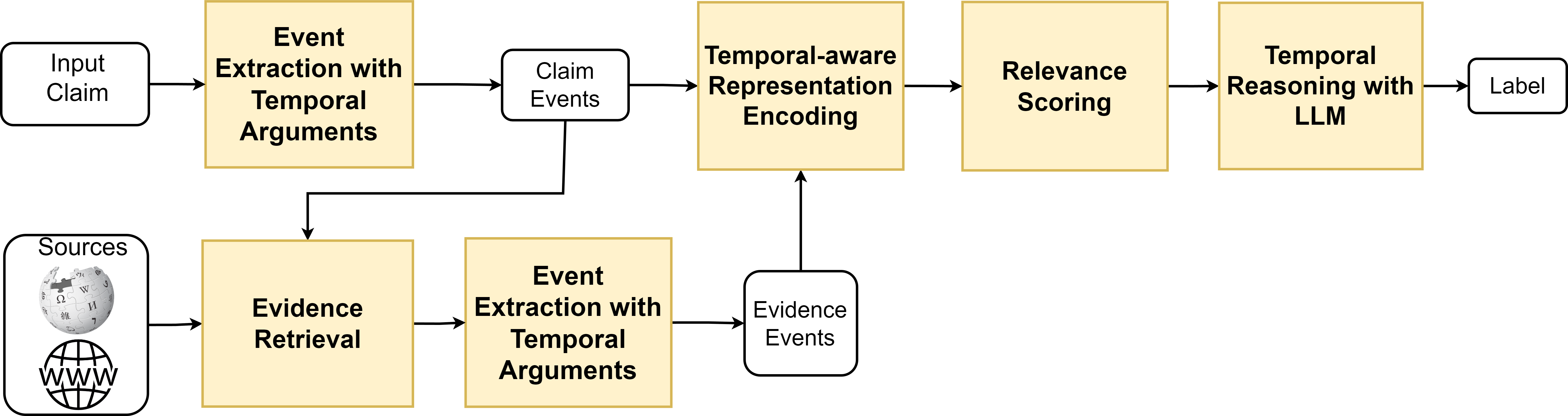}
\caption{Overview of TACV framework. }
\label{fig:tavc-overview}
\end{figure*}

We curate two datasets comprising of a diverse range of temporal claims which is crucial in evaluating the effectiveness of TACV.     
 We leverage the temporal reasoning ability of  large language model (LLM) by providing it with  the top-k evidence sentences retrieved by TACV as context to  determine the veracity of the claims.
 Experiment results on multiple benchmark datasets demonstrate that the proposed approach  surpasses
state-of-the-art methods, is robust and able to handle real-world claims.

\section{Related Work}

Research on evidence-based claim verification typically formulates the problem as a natural language inference task, and classifies whether the evidence sentences support or refute the claim \cite{domlin, bert-fact-checking}. 
 Recent works such as GEAR \cite{gear}, KGAT \cite{KGAT}  and GCAT \cite{CGAT} have employed graph attention networks to  propagate information between evidence sentences, thereby improving the model's reasoning ability. GEAR  uses a graph attention network to capture the semantic interaction between evidence sentences. KGAT introduces  kernels to measure the importance of the evidence  and conduct fine-grained evidence propagation.
  CGAT incorporates external knowledge to inject commonsense knowledge into the model. 

Several works have attempted to take into account temporal information for claim verification. \cite{neural-mt-temporal-claims} focuses on the verification on economic claims against time series sources which are in  tabular format.
A claim is translated into Datalog rules which are used to  query the tabular evidences to verify the correctness of the claim. 
This work only deals with structured SQL data and does not 
 handle evidences in natural language.

The work in \cite{time-aware-reranking} 
considers 
the published date of the claim and evidence sentences, and re-ranks 
the sentences based on the proximity of their published dates to that of the claim.  ITR \cite{implicit-temporal-reasoning} exploits the temporal proximity between the claim's publication date and evidence's publication date to create time representations for  temporal reasoning. 
These works do not consider temporal expressions in the claim/evidence sentences. However, such temporal metadata  may not be readily available. In contrast, our approach handles temporal expressions within the sentences themselves and do not require such external temporal metadata.

\section{Proposed Framework}


An overview of the proposed TACV framework is given in 
Figure~\ref{fig:tavc-overview}. 
Given a temporal claim, we extract claim events with their associated temporal expressions from the claim.
To obtain more information about the claim,
we use a sequence-to-sequence entity linking model GENRE \cite{genre} to retrieve documents from sources such as Wikipedia articles.
Each sentence from the retrieved documents is sent to the event extraction module to obtain evidence sentence events.
Next, we pair the extracted claim events with the evidence sentence events to create temporal-aware representations. This step facilitates the identification of the top-$k$ most relevant evidence sentences, which are deemed potentially useful for verifying the claim event. 
Utilizing the top-$k$ evidence sentences as context, the framework harnesses the temporal reasoning capabilities of Large Language Models (LLMs) to ascertain whether the evidence supports or refutes the claim event, or if the evidence is insufficient for verification.
Finally, these labels  are aggregated to obtain the final label for the input claim.

\subsection{Event Extraction with Temporal Arguments}

 In general, an event has two types of information: (a) core information 
 such as who is involved, what is happening, and where it is happening; 
 and (b) temporal expression which includes specific dates, time duration and event  ordering.
We employ an off-the-shelf Semantic Role Labeling (SRL)\footnote{https://demo.allennlp.org/semantic-role-labeling}  from AllenNLP \cite{allennlp-srl}
 to extract all the events mentioned in the claim or evidence sentences.
  Specifically, each sentence is fed into the SRL model, which outputs a list of predicates along with their arguments. Each predicate corresponds to an event. The core information comprises of the concatenation of phrases related to the predicate and non-temporal arguments.
  The temporal information comprises of the phrases related to the temporal arguments. We apply this process to the claim and evidence sentences to extract claim events and evidence events. 

  For instance, there are two predicates
  in the claim  \textit{"Henry Condell published his First Folio in 1623 and performed several plays for his career around October 1620."}. The first predicate
 \textit{"published"} has arguments \textit{"Henry Condell", "the First Folio"} along with the temporal argument \textit{"in 1623"}; while the second predicate
 \textit{"performed"} has the arguments \textit{"Henry Condell", "several plays"} as well as the temporal argument  \textit{"around October 1620"}.
 Similarly, for the 
 evidence sentence  \textit{"Henry Condell  ended his stage career in 1619."}, we have the 
  predicate \textit{"ended"} with arguments \textit{"Henry Condell", "stage career"} and temporal argument \textit{"in 1619"}.

The claim events and lists of extracted events in the evidence sentences
are paired before passing to a temporal-aware encoder.

\subsection{Temporal-aware Representation Encoding}

Our temporal-aware encoder utilizes the contextual and temporal information in the claim and evidence events to ensure that the time encoding is not just a static representation of time. This is in contrast to existing  work \cite{indcutive_temporal_representation} that uses series of sine and cosine for functional encoding of time element.
\begin{figure*}[t!] 
\centering
\includegraphics[width=0.67\textwidth]{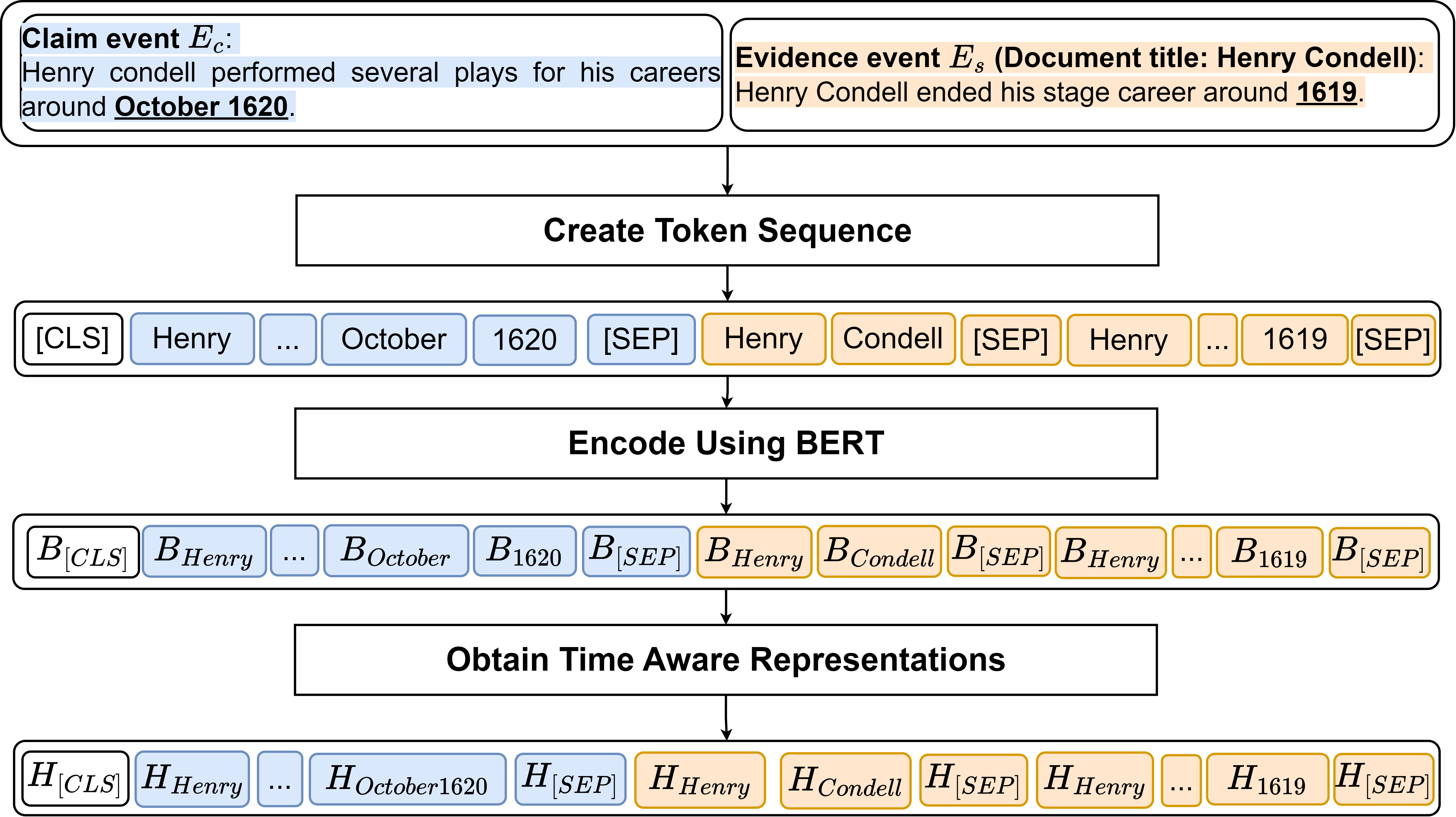}
\caption{Temporal-aware representation encoding.}
\label{fig:encoder-module}
\end{figure*}

Let $<$$E_c$, $E_s$$>$ be a <claim event, sentence event> pair.
We first create the sequence
$([CLS]+E_c+[SEP]+Title+[SEP]+E_s+[SEP])$
where 
$[CLS]$ is the special start token, $[SEP]$ is the separator token, and
$Title$ is the title of the document from which the sentence event $E_s$ is obtained. 
The sequence is then passed to BERT to obtain the contextual representation $B$ (see Figure~\ref{fig:encoder-module}).


Next, we apply mean pooling on the date tokens, followed by positional encoding \cite{transformer}. 
For example, consider the  temporal phrases \textit{October 1620}  and \textit{1619} in $E_c$  and $E_s$ respectively. The position $pos$ for \textit{1619} is  \textit{0}, while that for  \textit{October 1620} is \textit{21}, indicating that they are \textit{0} and \textit{21} months apart from the earliest date in the text (which is \textit{1619}). 
 Note that the temporal arguments could be of different granularities, and  the smallest granularity to used to compute $pos$. 
 
 Given the  $pos$ value, the  temporal encoding  is a vector of $d$ dimension, denoted as $TE_{pos}$, where the i$^{th}$ element is given by
\[ 
TE_{pos}[i] = \left\{
\begin{array}{ll}
      sin(\frac{pos}{10000^{i/d}}) & \mbox{if~} i \mbox{~is even} \\
      cos(\frac{pos}{10000^{(i-1)/d}}) & \mbox{otherwise} \\
\end{array} 
\right. 
\]

This temporal encoding  satisfies the following properties:

\smallskip
\noindent\textbf {1. Triangular inequality.} Given three distinct dates $t_1, t_2, t_3$ in chronological order with position  values $p_1, p_2, p_3$, the distance between the temporal encodings $TE_{p_1}$ and $TE_{p_2}$ is less than  the distance between $TE_{p_1}$ and $TE_{p_3}$.

\smallskip   
\noindent\textbf {2. Distance preservation.} Given the position  values $p_1, p_2, p_3, p_4$ of four distinct dates $t_1, t_2, t_3, t_4$, if $(p_2 - p_1) = (p_4 - p_3)$, then the distance between the temporal encodings  $TE_{p_2}$ and $TE_{p_1}$ is the same as the distance between $TE_{p_4}$ and $TE_{p_3}$.

\smallskip
We feed the temporal encodings to the transformer to obtain  the date representations
$\hat{B}$. 
The resulting temporal-aware representation of a <claim event, sentence event> pair is the sequence $R = (H_{[CLS]},$ $H_1, \cdots H_d)$
where
$$
H_j = \begin{cases}
B_j  & \text{if } j^{th} \text{ token is not a date}\\
\hat{B}_j & \text{if } j^{th} \text{ token is a date}
\end{cases}
$$
and 
 $H_{[CLS]}$ is the average pooling of  $H_j, 1 \leq j \leq d$.

\subsection{Relevance Scoring }

The temporal-aware representations obtained  are used to construct an event-level graph $G_{event}$.
Each node $i$ represents a <claim event, sentence event> pair and is initialized with its corresponding temporal-aware representation  $R^i$.
The nodes are fully connected to each other.
We utilize a Graph Attention Network (GAT)  to propagate information among the nodes in $G_{event}$.

Given two nodes $i$ and $j$,
we compute the token-level attention weight  $w^{i \rightarrow j}$ where the $p^{th}$ entry in $\mathbf{w}^{i \rightarrow j}$ is given by:
\begin{equation}
\mathbf{w}^{i \rightarrow j}[p] = \sum_{q} \mbox{sim}(R^i_p, R^j_q)
\label{eq:phrase-att-weight}
\end{equation}
where $sim$ is cosine similarity function and $R^j_q$ is the $q^{th}$ element in $R^j$.

We normalize $\mathbf{w}^{i \rightarrow j}$ through a  softmax function before applying this attention weight to the representation $R^i$. The information propagated  from node $i$ to node $j$ is given by:
\begin{equation}
    \mathbf{z}^{i \rightarrow j} = R^j_0 \circ (\mathbf{w}^{i \rightarrow j} \cdot R^i)
\end{equation}
where $R^j_0$ is the $[CLS]$ token in $R^j$ and $\circ$ denotes concatenation.

With this, we update the  representation of ${R}^j$ as follows:
\begin{equation}
     R^j = \sum_{i}\beta^{i \rightarrow j}\cdot \mathbf{z}^{i \rightarrow j}
\end{equation}
where $\beta^{i \rightarrow j}$ is the sentence-level attention weight from $i$ to $j$ computed 
 as follows:
\begin{equation}
\beta^{i \rightarrow j} = \textbf{W} \cdot (\mathbf{z}^{i \rightarrow j})^T
\end{equation}
where $\textbf{W} \in R^{1 \times 2d}$ denotes the weight matrix of a linear transformation and $(\mathbf{z}^{i \rightarrow j})^T$ is the transpose of $\mathbf{z}^{i \rightarrow j}$.



We compute the relevance score of each evidence sentence to a claim event by applying element-wise max operation \cite{gear} on the updated representations followed by  a linear layer.

\subsection{Temporal Reasoning with LLM}

Finally, we leverage the capabilities of LLM \textit{text-davinci-003} from OpenAI to perform temporal reasoning.
We design the following prompt to use the top-k relevant evidence sentences as context for LLM to reason and determine a label for each claim event.

\begin{table*}[t!]
\caption{Characteristics of temporal claim datasets. }
\vspace*{-0.1in}
\begin{center}

\begin{tabular}{|c|c|c|c|c|c|c|c|c|}
\hline
 &  \multicolumn{4}{c|}{T-FEVER} & \multicolumn{4}{|c|}{T-FEVEROUS} \\ \cline{2-9}
& \multicolumn{2}{c}{Single event} & \multicolumn{2}{|c|}{Multiple events}& \multicolumn{2}{c}{Single event} & \multicolumn{2}{|c|}{Multiple events}\\  \cline{2-9} 
 &   Train set  & Test set & Train set  & Test set &   Train set  & Test set & Train set  & Test set  \\ \hline
  Ordering & 20,625 & 2,805 &   1,009 & 161 &  17,546 & 1,910 & 39,402 & 4,175  \\\hline
  Duration & 456  & 75 &  21  & 3 &  374 & 51 & 729 & 106\\ \hline

 \end{tabular}
\end{center}
\label{table:statistics}
 \end{table*}

\textit{\small "You are provided with a claim  and evidence sentences. Perform text classification to determine whether the evidence support, refute, or do not enough information to verify the claim. The judgment based on the general aspect and the temporal aspect of the claim. Only base your decision on the information explicitly stated in the evidence. Please return the following information in JSON format: PREDICTED\_LABEL: Either 'SUPPORTS', 'REFUTES', or 'NOT ENOUGH INFO'. If the evidences does not address all key aspects of the claim or does not provide any information related to the claim then the returned label should be 'NOT ENOUGH INFO'. If the evidences contradicts general or temporal aspects of the claim, the returned label should be 'REFUTES'. If the evidences support all aspects of the claim, the returned label should be 'SUPPORTS'.}

The final label for a claim is determined as follows:
 If any event reveals factual discrepancies, the entire claim is deemed REFUTE. Conversely, if all events align with the  facts in the evidence sentences, the claim receives a SUPPORT label. In cases where certain events lack sufficient evidence while other events may be corroborated,  the overall verdict is NOT ENOUGH INFO.

  \begin{table*}[t!]
\caption{Characteristics of the experimental datasets. }
\vspace*{-0.1in}
\begin{center}
\begin{tabular}{|l|l|c|c|c|c|c|c|}
\hline
 & & \multicolumn{3}{c|}{Train Set} & \multicolumn{3}{|c|}{Test Set} \\ \cline{2-8}
Type & Dataset  &   Support & Refute & NEI  & Support & Refute & NEI \\ \hline

 Temporal claims & T-FEVER & 10,784 & 8,007 & 3,238 & 1,015 & 1,285 & 737 \\\cline{2-8}
  & T-FEVEROUS & 30,366 & 26,032 & 1,041 & 2,991 & 2,927 & 225  \\ \hline
General claims & FEVER & 80,035 & 29,775 & 35,639 & 3,333 & 3,333 & 3,333 \\ \cline{2-8}
 &FEVEROUS & 41,835 & 27,215 & 2,241 & 3,372 & 2,973 & 1,500 \\ \hline 
Real world  claims
 & LIAR & 1,683 & 1,998 &  -&  211 & 250 & -\\ \hline

 \end{tabular} 
\end{center}
\label{table:statistics2}
 \end{table*}
 
\section{Temporal Claim Datasets}

We derive two datasets for temporal fact verification from existing fact verification datasets FEVER \cite{fever}, FEVER2 \cite{fever2} and FEVEROUS \cite{feverous}. The original datasets comprise of synthetic general claims generated by modifying sentences from Wikipedia, and are labelled as SUPPORT, REFUTE, or NEI, along with their evidence sentences.
Although these datasets may have temporal claims, the verification of these claims is based on the general aspect instead of the temporal aspects. 
For example, the sample claim 
    "DSV Leoben, an Australian association football club which was founded in 1927" is managed by Austria Ivo Golz." is refuted 
    based on the ground truth evidence: 
    "DSV Leoben is an Austrian association football club based in Leoben." 
Here, we augment the dataset with new claims by manipulating
the temporal context of existing claims such as "DSV Leoben was founded in 1928".

The creation of the new datasets follows a three-stage process: 
 (1) detection of temporal claims, (2) tagging the types of  temporal expression in the claims, (3)  augmentation with synthetic claims.
We first identify temporal claims from the general fact verification datasets by
extracting claims with at least one temporal argument.
These claims are tagged according to their temporal expression type. This is achieved through the use of regular expression pattern matching to distinguish between the temporal expression types, namely 
  ordering (indicated by words such as \textit{"before"}, \textit{"after"}), and
  duration (phrases like \textit{"for 5 years"}, \textit{"over 3 months"}).
Claims that are not tagged are filtered out.

 
We also  augment the datasets with new claims
by manipulating the temporal context of existing claims, thereby creating a new scenario where either the new claim is no longer valid according to the altered date, or it remains valid in spite of the altered date. 
We start with original claims that are labeled as SUPPORT, and adjust the temporal argument such  that it is either disputed by the evidence sentences (leading to the new claim being labeled as REFUTE) or is in agreement with the  evidence sentences (which results in the new claim being labeled as SUPPORT). The evidence sentences are  the ground-truth evidence sentences provided by the original datasets.
Specifically, the following steps are taken to generate new REFUTE claims: 

\smallskip
\noindent \textbf{Ordering.}  We   extract the temporal predicates and  dates from the claim's temporal argument.
    \begin{itemize}
        \item 
    If the  temporal predicate is  \textit{"in", "on"} or \textit{"at"}, we replace the extracted claim date by
     adding or subtracting a random number to the date so that the new claim date is no longer supported by the date(s) in the evidence sentences. 

 \item   If the temporal predicate is \textit{"before"}, we identify the most recent date from the evidence sentences. Then we replace the predicate with  \textit{"after"} and adjust the claim date to the identified  date after adding a random number.
   Similarly, if the predicate is \textit{"after"}, we switch it to \textit{"before"} and revise the claim date to the earliest date mentioned in the 
    evidence sentences, again incremented by a random number.
   
\item   If the temporal predicate is \textit{"from"}, we find the most recent date  from the evidence sentences, and replace the claim date by the identified date after adding a random number.

\item If the temporal predicate is \textit{"between"} with two temporal arguments date$_1$ and date$_2$,  we add a random number to date$_2$ to get a new date$_3$. Then we replace  date$_1$ with date$_3$, and replace date$_2$ with date$_3$ after adding another random number. This ensures that the new range falls outside the original range. 

  \end{itemize}   
  
\noindent \textbf{Duration.} The temporal predicate is either \textit{"for", "over"}, or \textit{"within"}, accompanied by a temporal argument indicating the duration period. We adjust this  argument by randomly increasing or decreasing its value, thereby creating a new duration that diverges from the original context.
  \smallskip

Likewise, we augment the datasets with new claims that are labeled as "SUPPORT" by ensuring that the modified temporal arguments remain consistent with the evidence sentences.
The dataset derived from the FEVER and the FEVER2.0 datasets is called \textbf{T-FEVER}, while the dataset that is derived FEVEROUS is called \textbf{T-FEVEROUS}. 
 Each claim in our datasets can involve either a single event or multiple events. These events are characterized by their temporal expressions focusing on two key aspects: ordering and duration. A detailed breakdown of these characteristics for each dataset is provided in  \autoref{table:statistics}. 
These datasets will be made available on Github.

We evaluated the quality of our new datasets by randomly sampling 300 claims each from T-FEVER and T-FEVEROUS. Two human assessors, equipped with the
necessary background and skills,
were tasked  to determine the accuracy of a claim's label by referencing the ground truth evidence sentences. 
Our findings indicate that 97\% of the claims in T-FEVER and 98\%  in 
 T-FEVEROUS have the correct labels. The remaining claims are wrongly labelled  as SUPPORT/REFUTE when they should be labelled as NOT ENOUGH INFO.
 For example, consider the claim “Ashley Graham was on a magazine cover in 2018.” and the corresponding evidence sentence “In 2017, Graham became the first plus-size model to appear on the covers of British and American Vogue.”. The claim was incorrectly labelled as  REFUTE when it should be 
NOT ENOUGH INFO. This is because even if Graham was on the magazine cover in 2017, this does not imply that she cannot appear on the cover in 2018.


\smallskip
\noindent\textbf{Training.}
We use these two datasets in the training of the temporal-aware encoder in TACV. We employ triplet loss to learn representations that  maximize the margin between relevant and non-relevant sentences to a specific claim event. The training data is composed of 
triplets, each comprising of a claim event, a relevant sentence from the ground truth evidence, and a non-relevant sentence selected from Wikipedia articles.
Since the ground truth evidence sentences pertain to a claim as a whole rather than to specific events in the claim, we engage two human annotators to identify which  evidence sentences  are relevant  to  individual claim event. This enables a more precise training for the TACV's temporal-aware encoder.

\section{Performance Study}
In this section, we present the results of  experiments to evaluate the effectiveness of the TACV framework for temporal fact verification.
We show that TACV performs well not only on the new temporal  T-FEVER and T-FEVEROUS datasets, but also on the standard benchmark FEVER and FEVEROUS datasets as well as the real world LIAR dataset \cite{liar}.
 \autoref{table:statistics2} gives the details of these  datasets. 

We use label accuracy and FEVER score as  the evaluation metrics. Label accuracy measures the proportion of correct predictions made by the model out of all predictions. This metric ignores whether the evidence sentences directly contribute to the prediction.
In contrast, 
FEVER score only marks a prediction as correct if the predicted label  is correct and the retrieved evidence directly contributes to the determination of the label.

TACV uses  Huggingface's implementation of BERT$_{base}$, a  pre-trained language model, to encode the tokens in the claim events and evidence sentence events.   For the temporal-aware representation encoding, a transformer with 
 two layers and eight heads, having a  dimension of 768, is used.
The training is conducted over five epochs with  a batch size of 8, and learning rate of 5e-6. We apply the AdamW \cite{adamw} optimizer with  a fixed weight decay and select the best performing model for evaluation on the test set. 

\begin{figure}[t!] 
\centering
  \centering
  \includegraphics[height= 1.2in,width=.22\textwidth]{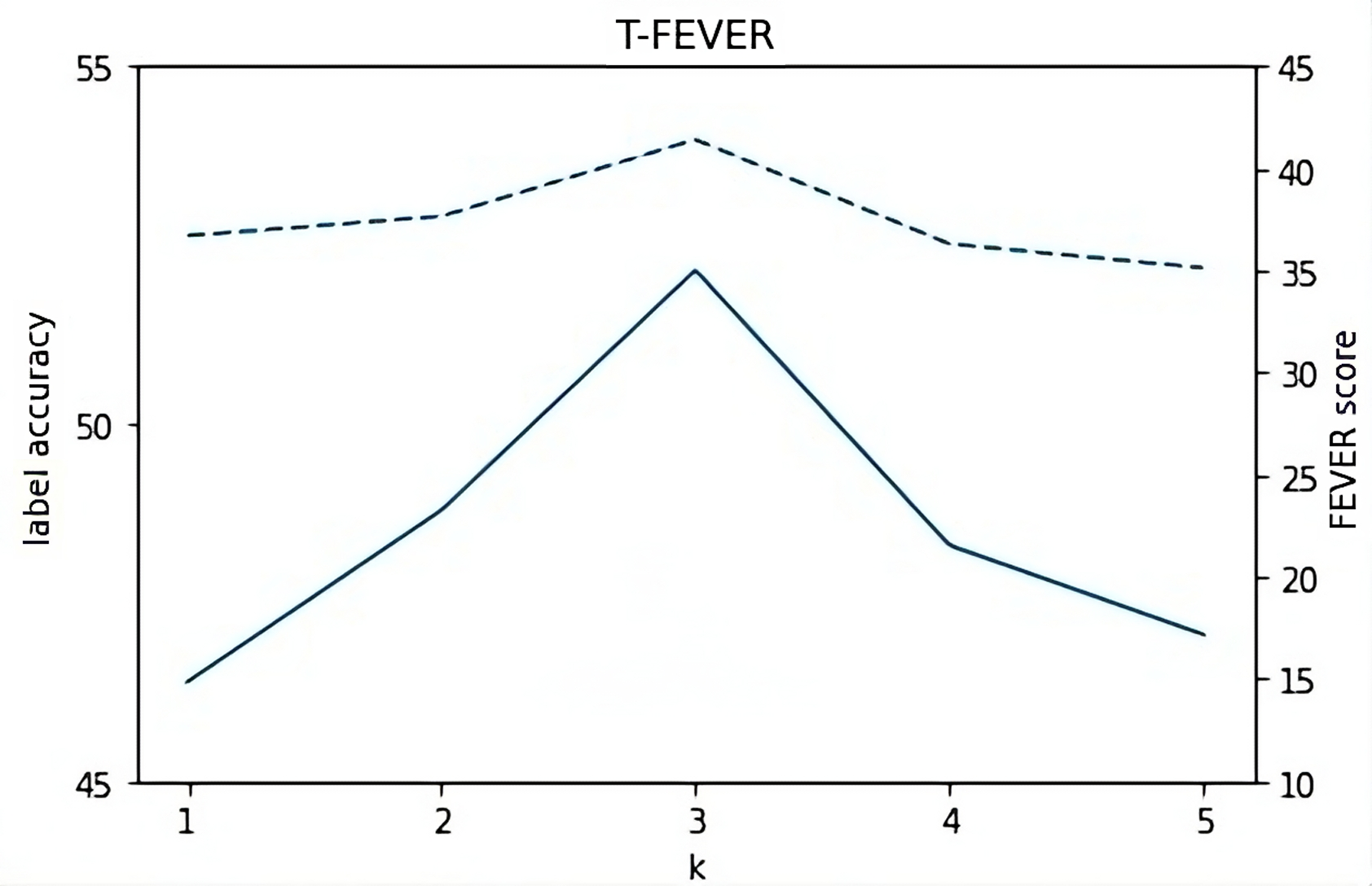}
  \includegraphics[height= 1.2in,width=.22\textwidth]{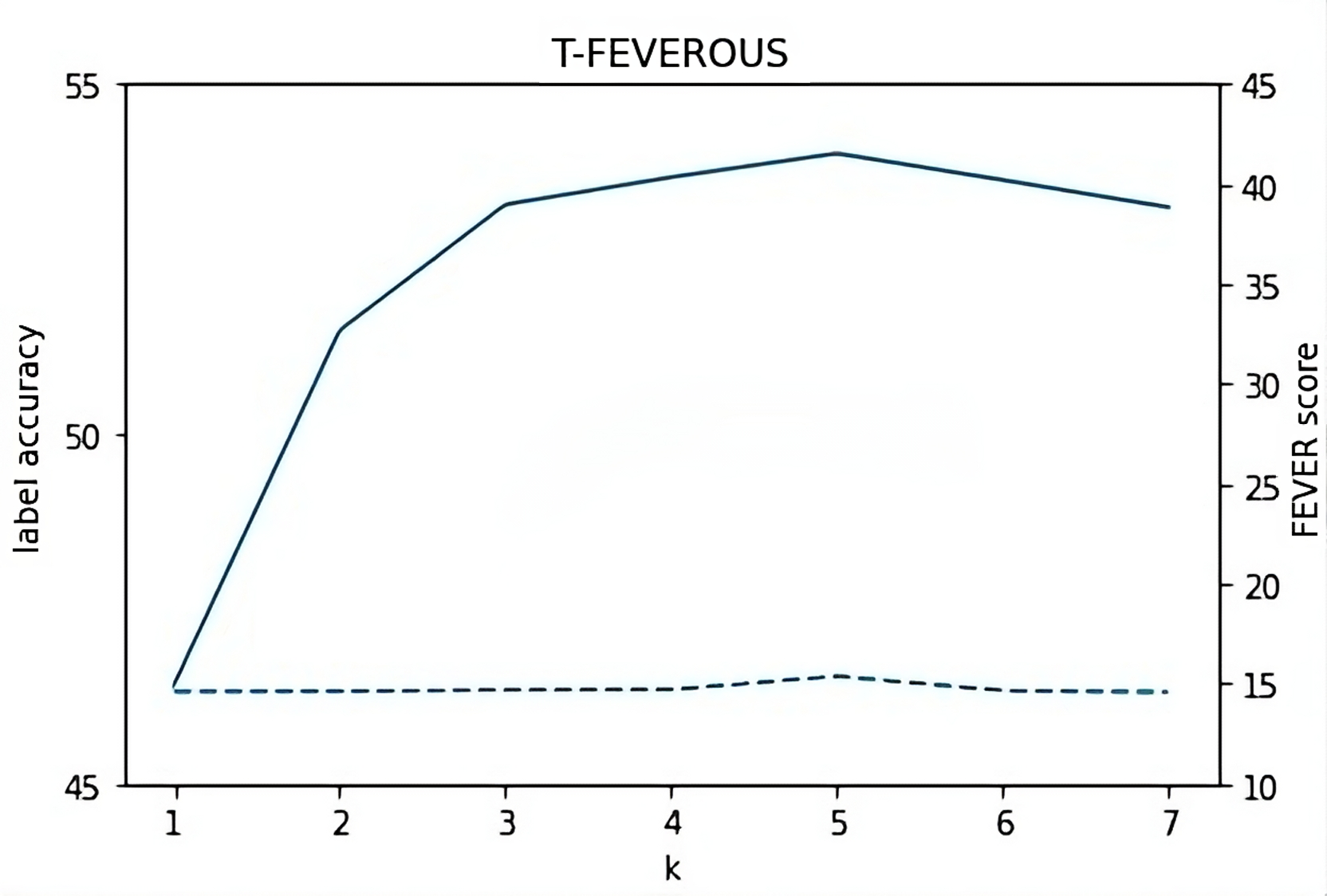}

\caption{Effect on $k$ on TACV. Solid line depict label accuracy, dotted line depict FEVER score.}
\label{fig:scores-k-variation}
\end{figure}

\subsection{Sensitivity Experiments}

We first conduct  sensitivity experiments to examine the performance of TACV as we vary the number of top-$k$ relevant evidence sentences for temporal reasoning. 
 Figure~\ref{fig:scores-k-variation} shows the label accuracy and FEVER score for different $k$ values on the T-FEVER and T-FEVEROUS datasets.
We see that the optimal performance is attained when $k= 3$ for T-FEVER, and $k=5$ for T-FEVEROUS.
As such, we use the top-3 sentences in T-FEVER, and the top-5 sentences in T-FEVEROUS with the highest relevance scores to form the context for the LLM to output the label of each claim event.

\begin{table*}[t!]
\caption{Results of comparative study on temporal claims.}
\vspace*{-0.1in}
\begin{center}
\begin{tabular}{|c|cc|cc|}
\hline
& \multicolumn{2}{c|}{T-FEVER}  & \multicolumn{2}{c|}{T-FEVEROUS}  \\\hline
Methods & {Label acc.} &{FEVER score} &{Label acc.} & {FEVER score}  \\ \hline
KGAT & 44.28 & 33.61 & 15.69 & 4.59 \\\hline
CGAT & 44.38 & 33.91 & 16.58 & 4.29  \\\hline
ITR & 44.05 & 30.88 & 31.66&8.63  \\\hline
TACV & \textbf{52.15} & \textbf{41.42} & \textbf{54.01} & \textbf{15.38}   \\
\hline
\end{tabular}\\
\label{table:comparative}
\end{center}
\end{table*}


\begin{table*}[t!]
\caption{Detailed results on T-FEVER.}
\vspace*{-0.1in}
\begin{center}
\begin{tabular}{|c|c|c|c|c|c|c|c|c|}
\hline
 & \multicolumn{4}{c|}{Single Event} & \multicolumn{4}{c|}{Multiple Event} \\
\cline{2-9} &\multicolumn{2}{c|}{Ordering} & \multicolumn{2}{c|}{Duration} & \multicolumn{2}{c|}{Ordering} & \multicolumn{2}{c|}{Duration}  \\\hline
& Label acc. & Fever score & Label acc. & Fever score & Label acc. & Fever score & Label acc. & Fever score \\ \hline
KGAT & 44.4 & 34.30 & 36.36 &34.84 & 45.2 & 31.60 & 41.67 &41.67 \\\hline
CGAT & 44.6 & 34.14 & 37.87 & 34.84 & 43.4 & 30.60 &  41.67 & 41.67 \\\hline
ITR & 44.48 & 31.63 & 45.45 & 39.39 & 42.22 & 26.2 & 33.33 & 33.33 \\\hline
TACV & \textbf{52.14}  & \textbf{41.56} &\textbf{50.00}   & \textbf{46.96} & \textbf{52.60}  &\textbf{40.00} & \textbf{58.33} & \textbf{58.33}    \\\hline
\end{tabular}
\label{table:tfever-results}
\end{center}
\end{table*}

\begin{table*}[t!]
\caption{Detailed results on T-FEVEROUS.}
\vspace*{-0.1in}
\begin{center}
\begin{tabular}{|c|c|c|c|c|c|c|c|c|}
\hline
 & \multicolumn{4}{c|}{Single Event} & \multicolumn{4}{c|}{Multiple Event} \\
\cline{2-9} &\multicolumn{2}{c|}{Ordering} & \multicolumn{2}{c|}{Duration} & \multicolumn{2}{c|}{Ordering} & \multicolumn{2}{c|}{Duration}  \\\hline
& Label acc. & Fever score & Label acc. & Fever score & Label acc. & Fever score & Label acc. & Fever score \\ \hline
KGAT & 20.68 & 4.41 & 8.69 & 0 & 15.73 & 4.26 & 17.16 & 3.73 \\\hline
CGAT & 18.57 & 4.61 & 8.69 & 0 & 15.91 & 4.59 & 18.65 & 3.73 \\\hline
ITR & 31.32 & 5.72 & 17.39 & 4.34 & 32.02 & 9.21 &  28.35 & 5.97 \\\hline
TACV & \textbf{50.70} &\textbf{11.14} & \textbf{43.47} & \textbf{26.08} & \textbf{54.72} & \textbf{18.53} & \textbf{47.01} & \textbf{16.41}  \\\hline
\end{tabular}
\label{table:tfeverous-results}
\end{center}
\end{table*}

\begin{table*}[t!]
\caption{Results of comparative study on general and real world claims.}
\vspace*{-0.1in}
\begin{center}
\begin{tabular}{|c|cc|cc|c|c|}
\hline
& \multicolumn{2}{c|}{FEVER} & \multicolumn{2}{c|}{FEVEROUS} & LIAR & T-LIAR \\\hline
Methods &  {Label acc.} &{FEVER score}& {Label acc.} & {FEVER score}&{Label acc.} &{Label acc.}\\ \hline
KGAT &  74.07 &70.38 & 34.94 & 11.25& 46.20 & 69.44\\\hline
CGAT & 76.39 & 73.15 & 39.70 & 12.52 & 45.77 &72.22 \\\hline
ITR & 73.36 & 70.04 & 44.20&14.39 & 49.24 &69.44\\\hline
TACV & \textbf{76.42} &\textbf{73.16} &\textbf{53.97} & \textbf{15.08} &\textbf{62.86} &\textbf{83.33}\\\hline
\end{tabular}\\
\label{table:comparative2}
\end{center}
\end{table*}

\subsection{Comparative Experiments}

In this  set of experiments, we compare  TACV with the following state-of-the-art evidence-based fact verification baselines:
\begin{itemize}
    \item KGAT \cite{KGAT}. This method employs transformer-based architecture 
to obtain claim-sentence pair representations before utilizing
a fine-grained Kernel graph attention networks (GAT) to aggregate the evidence for the
claim verification. It employs Athene \cite{athene} to retrieve  evidence sentences. 

    \item CGAT \cite{CGAT}. This method incorporates  external knowledge from ConceptNet to enrich the contextual representations of evidence sentences before employing GAT  to propagate the information among the evidence sentences to verify the claim veracity. Same as KGAT, evidence sentences are obtained using Athene.

  \item ITR  \cite{implicit-temporal-reasoning}. This method assumes evidence sentences are given as input and focuses on using them for temporal reasoning.
 For fair comparison, we use the evidence sentences retrieved by our TACV as input to ITR.
  
\end{itemize}

\noindent\textbf{Results on temporal claims.}
Table~\ref{table:comparative} shows the label accuracy and FEVER score of the various methods on the new
temporal claim datasets T-FEVER and T-FEVEROUS. 
We see that TACV outperforms existing methods by a large margin.

Tables~\ref{table:tfever-results} and \ref{table:tfeverous-results} provide further analysis of the results on the new temporal claim datasets. We breakdown the results based on whether a claim consists of single events or multiple events, as well as the type of temporal expressions such as ordering or duration. 
We see that TACV consistently outperforms state-of-the-art methods in terms of both label accuracy and FEVER score for claims involving single and multiple events.
For claims with multiple events, TACV shows a larger margin of improvement indicating the importance of analyzing claims at the event level.
Further, there is a significant improvement in the FEVER score implying that TACV  is able to retrieve evidence sentences relevant to the claim.

\smallskip
\noindent\textbf{Results on general and real world claims.}
We also validate the ability of TACV to handle both the original synthetic general claims in FEVER and FEVEROUS, as well as  real-world claims in LIAR which comprises of  statements compiled from PolitiFact.com. For each claim in the LIAR dataset, we feed the claim sentence into BING search to retrieve the top-2 articles and all the sentences from these articles are used as potential evidence sentences.
In addition, we  identify 363 temporal claims (209 SUPPORT and 154 REFUTE) in the LIAR to create a T-LIAR dataset.

\begin{table*}[t!]
\caption{Results of Ablation Studies.}
\vspace*{-0.1in}
\begin{center}
\begin{tabular}{|c|cc|cc|}
\hline
& \multicolumn{2}{c|}{T-FEVER}  & \multicolumn{2}{c|}{T-FEVEROUS}  \\\hline
Methods & {Label acc.} &{FEVER score} &{Label acc.} & {FEVER score}  \\ \hline
TACV w/o event extraction & 46.92 & 39.47 & 39.64 & 12.02 \\\hline
TACV w/o  Temporal-aware encoder & 49.22 & 38.88 & 52.14 & 13.07  \\\hline 
TACV w/o GAT & 50.60 & 40.07 & 52.84 & 13.91  \\\hline
TACV & \textbf{52.15} & \textbf{41.42} & \textbf{54.01} & \textbf{15.38}   \\\hline
TACV (GPT4) & \textbf{55.08} & \textbf{42.17} & \textbf{56.56} & \textbf{18.98}   \\
\hline

\end{tabular}\\
\label{table:ablation-studies}
\end{center}
\end{table*}

\begin{table*}[t!]
\caption{Sample Claim from T-FEVER.}
\vspace*{-0.15in}
\begin{center}
\small
\begin{tabular}{|p{0.05\textwidth} | p{0.18\textwidth} | p{0.50\textwidth}|p{0.08\textwidth}|p{0.08\textwidth}|}
\hline
\multicolumn{5}{|l|}{Claim: Osamu Tezuka practiced drawing as a young child in 2000. \hfill Ground Truth Label: REFUTE} \\ \hline
Method& Events & Retrieved Sentences &  Event Label & Claim Label \\ \hline
TACV & \textbullet{ Osamu Tezuka \textcolor{blue}{practiced} drawing as a young child \textcolor{blue}{in 2000}.} & 
\textbullet{ As a young child Tezuka began to practice drawing so much that his mother would have to erase pages in his notebook in order to keep up with his output. } \newline 
\textbf {\textbullet { Tezuka died of stomach cancer in 1989. }} \newline \textbullet{ Born in Osaka Prefecture, his prolific output, pioneering techniques, and innovative redefinitions of genres earned him such titles as "the father of manga", "the godfather of manga" and "the god of manga".}   & REFUTE & REFUTE \\ \hline
CGAT  & -
&\textbullet{ His legendary output would spawn some of the most influential , successful and well received manga series including Astro Boy , Jungle Emperor Leo , Black Jack , and Phoenix , all of which would win several awards .} \newline \textbullet{ As a young child Tezuka began to practice drawing so much that his mother would have to erase pages in his notebook in order to keep up with his output .} \newline \textbullet{ In elementary school he would develop his own comics and illustrations from which his skills would become formidable .}  & - & NEI \\ \hline

\end{tabular}
\label{table:case-studies-fever}
\end{center}
\end{table*}

Table~\ref{table:comparative2} shows the results. We see that 
TACV remains robust in the original FEVER and FEVEROUS datasets
and can generalize well to real world LIAR dataset.  Moreover, TACV has a big lead in the label accuracy of the real world temporal claim T-LIAR dataset, raising the confidence that TACV can be used for the verification of temporal claims in real world settings.

\subsection{Ablation Studies}

We implement the following variants of TACV  to examine the effect of the key components in TACV:
\begin{enumerate}
\item TACV without event extraction. Instead of extracting events from claim and evidence sentences, we pass them directly to the temporal-aware representation encoding component. The top-k relevant sentences for each claim are passed to the LLM  to obtain the claim's label.

\item TACV without temporal-aware representation encoding. For this variant, we use BERT to obtain the encoding for  each pair of claim event and sentence event and use this representation for relevance scoring.

\item TACV without GAT. Here, we do not construct the $G_{event}$ graphs. Instead, we perform mean pooling over the token representations of the <claim event, sentence event> pairs. 

\item TACV (GPT4). Here, we also experimented with a better LLM by using GPT4-turbo.
 \end{enumerate}

\autoref{table:ablation-studies} shows the results.  The largest drop in both label accuracy and FEVER score occur when events are not  extracted  from claim and evidence. This is followed by the variant where temporal-aware representation encoding is not utilized. 
This suggests that identifying events in claims and evidence enhances the retrieval of relevant sentences for the subsequent claim verification process.
Also, using a better LLM  further improves the performance. 



\section{Case Studies}

\autoref{table:case-studies-fever} shows a claim from T-FEVER and the top-ranked  evidence sentences retrieved by each method. 
 The claim "Osamu Tezuka practiced drawing in 2000" has a single event "practiced" with  temporal argument "in 2000". 
TACV with its temporal-aware representation is able to retrieve the  sentence "Tezuka died of stomach cancer in 1989", which is also the ground truth evidence sentence (depicted in bold). With this, LLM concludes that the event label is REFUTE, and the overall claim label is REFUTE.
On the other hand, CGAT  retrieves evidence sentences that describe Tezuka's drawing practice but not the date, resulting in the NEI prediction.

\begin{table*}[t!]
\caption{Sample Claims from T-FEVEROUS. }
\vspace*{-0.15in}
\begin{center}
\small
\begin{tabular}{|p{0.05\textwidth} | p{0.18\textwidth} | p{0.5\textwidth}|p{0.08\textwidth}|p{0.08\textwidth}|}
\hline

\multicolumn{5}{|p{0.97\textwidth}|}{Claim: McDonaugh was appointed to the first managerial job on the 10th April 2019, and then he was awarded SPFL League 2 Manager of the Month, in December 2019. \hfill Ground Truth Label: REFUTE} \\ \hline
Method &  Events & Retrieved Sentences &  Event Label & Claim Label \\ \hline

\multirow{2}{0.18\textwidth}{TACV} 
& \textbullet { McDonaugh was \textcolor{blue}{appointed} to the first managerial job on the \textcolor{blue}{10th April 2019}. } & \textbf{\textbullet { McDonaugh was appointed to first managerial job succeeding Gary Jardine at Edinburgh City on 10 October 2017.}} \newline \textbullet { McDonaugh again won the SPFL League 2 Manager of the Month award in December 2019, winning all four games and keeping three clean sheets.} \newline \textbullet { He was awarded SPFL League 2 Manager of the Month in September 2018.} & REFUTE & REFUTE \\ \cline{3-3}
& \textbullet { McDonaugh was \textcolor{purple}{awarded} SPFL League 2 Manager of the Month, \textcolor{purple}{in December 2019}.} & \textbf{\textbullet { McDonaugh again won the SPFL League 2 Manager of the Month award in December 2019, winning all four games and keeping three clean sheets.}} \newline \textbullet { He was awarded SPFL League 2 Manager of the Month in September 2018.} \newline \textbullet { McDonaugh was appointed to first managerial job succeeding Gary Jardine at Edinburgh City on 10 October 2017.} & SUPPORT & \\ \hline

\multirow{1}{0.18\textwidth}{CGAT} 
& - & \textbullet{ James McDonaugh is a Scottish football manager, who is currently manager of Scottish League Two club Edinburgh City and a current UEFA Pro Licence holder.} \newline \textbf{\textbullet{ McDonaugh again won the SPFL League 2 Manager of the Month award in December 2019, winning all four games and keeping three clean sheets.}} \newline \textbullet{ He was awarded SPFL League 2 Manager of the Month in Sept 2018.}  & - & NEI \\ \cline{3-3}

\hline

\end{tabular}
\label{table:case-studies-feverous}
\end{center}
\end{table*}

\begin{table*}[t!]
\caption{Sample Claims from T-Liar. }
\vspace*{-0.15in}
\begin{center}
\small
\begin{tabular}{|p{0.05\textwidth} | p{0.18\textwidth} | p{0.5\textwidth}|p{0.08\textwidth}|p{0.08\textwidth}|}
\hline
\multicolumn{5}{|p{0.97\textwidth}|}{Claim: Illinois {suffered} 1,652 overdose deaths {in 2014}, of which 40 percent were associated with heroin and Illinois is {ranked} number one in the nation for a decline in treatment capacity{between 2007 and 2012}. \hfill Ground Truth: SUPPORT} \\ \hline
Method &  Events & Retrieved Sentences &  Event Label & Claim Label \\ \hline

{TACV} 
& \textbullet{ Illinois \textcolor{blue}{suffered} 1,652 overdose deaths \textcolor{blue}{in 2014} , of which 40 percent were associated with heroin }& \textbf{\textbullet{Illinois suffered 1,652 overdose deaths in 2014 – a 30 percent increase over 2010 – of which 40 percent were associated with heroin}} \newline \textbullet {Durbin claims 40 percent of drug overdose deaths in Illinois involve heroin} \newline \textbullet{However, the Illinois Department of Public Health, which reports preliminary and final drug overdose deaths to the CDC, puts the 2010 total at 1,284 and 1,700 in 2014 -- a slight discrepancy but not unusual when reporting overdose deaths as they often get revised} & SUPPORT & SUPPORT \\ \cline{3-3}
&\textbullet{ Illinois \textcolor{purple}{ranked} number one in the nation for a decline in treatment capacity \textcolor{purple}{between 2007 and 2012}.} & \textbf{\textbullet{A report published in August 2015 by the Illinois Consortium on Drug Policy at Roosevelt University, or ICDP, shows state-funded treatment capacity in Illinois fell by 52 percent from 2007-2012, the largest decrease in the nation}} \newline \textbullet{In 2007, Illinois ranked 28th in state-funded treatment capacity before dropping to No. 44, or third worst in 2012, behind Tennessee and Texas, respectively, according to the report.} \newline \textbullet {Durbin, who used statistics from this study, is correct when he says Illinois led the nation in the decline for state-funded treatment capacity.} & SUPPORT & \\ \hline

{CGAT} 
&- & \textbf{\textbullet{Illinois suffered 1,652 overdose deaths in 2014 – a 30 percent increase over 2010 – of which 40 percent were associated with heroin}} \newline \textbullet{ As for the other figures, the percent increase from 2010 is slightly more than 32 percent, and drug overdose deaths in 2014 that were associated with heroin is about 42 percent} \newline \textbullet{In 2007, Illinois ranked 28th in state-funded treatment capacity before dropping to No. 44, or third worst in 2012, behind Tennessee and Texas, respectively, according to the report}  & - & REFUTE \\ \cline{2-4}

\hline
\end{tabular}
\label{table:case-studies-liar}
\end{center}
\end{table*}

\autoref{table:case-studies-feverous} shows a claim from T-FEVEROUS. The claim  has two events
"appointed" (in blue) and "awarded" (in red) with temporal arguments "on the 10th April 2019" and "in December
2019" respectively. 
By decomposing the claim into events and their temporal arguments,  TACV is able to retrieve both ground truth sentences, one supporting the "awarded" event and the other contradicting the "appointed" event.   LLM  predicts the label REFUTE for the first event and SUPPORT for the second event. As such,  the  claim label is  REFUTE.
In contrast,  CGAT does not retrieve the evidence sentence regarding the job appointment and predicts the claim  as NEI.  

\autoref{table:case-studies-liar} shows a sample claim from T-Liar. 
The claim consists of two events: "suffered" (in blue) and "ranked" (in red), along with their temporal arguments "in 2014" and "between 2007 and 2012".
By breaking down the claim into events, 
TACV is able to retrieve sentences that confirm the date of  overdose deaths for the first event, and sentences that mention the period when  Illinois is ranked number one for decline in treatment capacity. This allows  LLM to  verify each event as  SUPPORT, and  TACV to correctly conclude that the overall claim  label as  SUPPORT.
On the other hand,
CGAT fails to  retrieve  sentences that reference the date when Illinois was ranked first for declined treatment capacity, leading to an incorrect prediction of the claim's label.

\section{Conclusion}

In this work, we have introduced the TACV framework  for temporal claim verification. By decomposing complex sentences into event-level units and incorporating the temporal proximity of events, TACV learns a temporal-aware representations for evidence retrieval. 
Extensive experiments on both synthetic and real world datasets 
show marked improvements in label accuracy and FEVER score over state-of-the-art. Case studies highlight  TACV's ability to retrieve relevant evidence  for the verification of temporal  claims.
Moreover, the  curated datasets T-FEVER and T-FEVEROUS provide a standardized benchmark and   resource for research in  temporal claim verification.
While TACV  marks a significant step forward in temporal claim verification, further research is needed  to handle highly convoluted sentences with implicit temporal references.

\begin{acks}
This work is supported by the Ministry of Education, Singapore, under its MOE AcRF TIER 3 Grant (MOE-MOET32022-0001). 
We would like to thank Svetlana Churina for her help in the annotation of the T-FEVER and T-FEVEROUS datasets.
\end{acks}

\clearpage
\bibliographystyle{ACM-Reference-Format}
\bibliography{factcheck}










\end{document}